\documentclass[conference]{IEEEtran}
\IEEEoverridecommandlockouts
\usepackage{cite}
 \usepackage{fancyhdr}
\usepackage{amsmath,amssymb,amsfonts}
\usepackage{algorithm}
\usepackage{algpseudocode}
\usepackage{graphicx}
\usepackage{textcomp}
\usepackage{xcolor}

\def\BibTeX{{\rm B\kern-.05em{\sc i\kern-.025em b}\kern-.08em
    T\kern-.1667em\lower.7ex\hbox{E}\kern-.125emX}}
\begin{document}

\title{Investigating ET-SoC-1 Vector and \\ Parallel Performance with sGEMM\\
}

\author{\IEEEauthorblockN{Richard Neale}
\IEEEauthorblockA{\textit{EPCC} \\
\textit{University of Edinburgh}\\
Edinburgh, UK \\
s2269111@ed.ac.uk}
\and
\IEEEauthorblockN{Nick Brown}
\IEEEauthorblockA{\textit{EPCC} \\
\textit{University of Edinburgh}\\
Edinburgh, UK \\
n.brown@epcc.ed.ac.uk}
}
\maketitle


\begin{abstract}
The Ainekko (formerly Esperanto) ET-SoC-1 is a massively parallel RISC-V accelerator. Its compute array comprises 1024 dual-threaded in-order ET-Minion cores (2048 harts) operating at 600~MHz, each core providing an eight-lane 256-bit packed-single (PS) SIMD datapath shared by its two harts, with access to LPDDR4X memory at a nominal 133~GiB/s. This paper presents a performance investigation using single-precision general matrix multiplication (sGEMM) as the benchmark kernel. Two device kernels are implemented and evaluated on the ET-SoC-1, these are a register-blocked PS SIMD baseline with two-step software pipelining, and a TensorFMA32 kernel that drives the same FMA datapath through the per-core tensor micro-operation sequencer, with double-buffered \texttt{A} scratchpad loading. Because both kernels are bounded by the same arithmetic peak, the comparison isolates the effect of operand delivery from that of arithmetic width. Performance is compared against CPU references (SpacemiT K3, Intel i5-11600, AMD Ryzen~9 9950X and the Arm Cortex-X925/A725 CPU of an NVIDIA DGX Spark) and two GPUs (NVIDIA GB10 Blackwell and AMD RDNA~4 RX~9060~XT).
\end{abstract}



\section{Introduction}
The ET-SoC-1, originally developed by Esperanto Technologies and now the technology further developed and maintained by Ainekko, is a manycore RISC-V accelerator built around a large array of simple dual-threaded in-order cores whose collective throughput is intended to exceed that of conventional accelerator designs whilst retaining programmability in a standard ISA~\cite{ditzel2022hardware,esperanto_etsoc1_datasheet}. The device includes both a packed-single (PS) SIMD extension, albeit not compliant with the RISC-V Vector extension~\cite{rvv_spec_2021}, providing eight-wide single-precision 32-bit floating-point (fp32) vector operations on a natively 256-bit floating-point register file, and a tensor micro-operation sequencer which drives that same eight-lane Fused Multiply Add (FMA) datapath to compute matrix products, producing a $16 \times 16$ fp32 tile over 512 cycles. 

General Matrix Multiplication (GEMM) is a dense linear kernel which is
a dominant workload in machine learning inference and training, as well as being the core of many scientific computing algorithms. Furthermore, it provides a structure which is simple enough to enable detailed performance analysis yet rich enough to highlight the differences between arithmetic throughput, memory hierarchy and parallel  execution~\cite{goto2008anatomy,vanzee2015blis,brown2025risc}. In this paper we explore single precision GEMM, sGEMM, on the ET-SoC-1 RISC-V accelerator. sGEMM is an ideal test algorithm with which to evaluate the ET-SoC-1's architectural capabilities and limitations which we highlight and describe. The principal contributions of this work are:
\begin{enumerate}
  \item A hand-written register-blocked PS SIMD sGEMM kernel for the ET-SoC-1, establishing a reproducible performance baseline.
  \item A TensorFMA32 kernel with double-buffered \texttt{A} scratchpad loading to overlap the tensor unit's compute with memory transfers.
  \item CPU SIMD (RVV 1.0, AVX512 and SVE2) and GPU (CUDA and ROCm) for performance comparisions, together with OpenBLAS sGEMM for accuracy checks.
\end{enumerate}


\section{Background}

\subsection{The ET-SoC-1}

\begin{figure}
    \centering
    \includegraphics[width=1\linewidth]{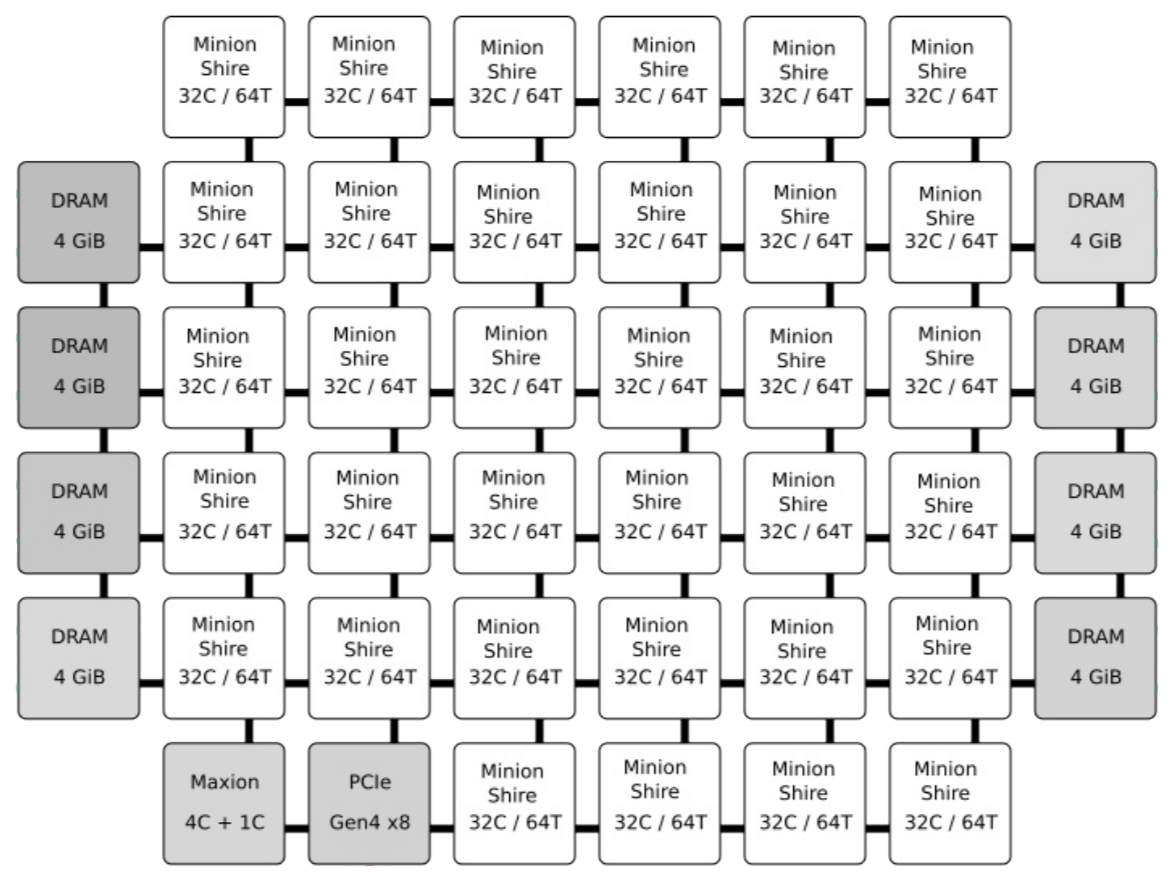}
    \caption{ET-SoC-1 Schema}
    \label{fig:et-soc-1}
\end{figure}

The ET-SoC-1 is a 7~nm FinFET device containing a total of 1093 RISC-V cores (\textit{figure \ref{fig:et-soc-1}}). This comprises 1088 in-order ET-Minion cores for compute, four out-of-order ET-Maxion cores for management tasks, and one simplified ET-Minion acting as a service processor. Cores are organised into a four-level hierarchy scaling down from chip, which contains shires, which themselves contain neighbourhoods which each contain cores~\cite{esperanto_etsoc1_datasheet,ditzel2022hardware,brown2025risc2}.

Each ET-Minion is a dual-threaded, in-order, single-issue RV64IMFC~\cite{waterman2016risc} core with a proprietary eight-lane SIMD vector/tensor unit and a private 4 KiB L1 data cache. The L1 cache is four-way set-associative with 64-byte cache lines, hence providing only 16 sets. Eight Minions form a \emph{neighbourhood}, and these share a 32~KiB L1 instruction cache (fronted by two 16-entry fully associative L0 microcaches, one per four cores), two page-table walkers and one set of performance counters. Four neighbourhoods and 4~MiB of shared SRAM form a \emph{shire}. That 4~MiB is built from four 1~MiB banks connected to a 512-bit crossbar, and each bank is independently configurable by software into different modes. This memory can act as shire-private L2 cache, as a slice of a chip-wide distributed L3 cache, or as globally addressable scratchpad memory. There is no separate, dedicated L3 array unless the SRAM is configured into that specific mode.

An ET-SoC-1 chip contains 34 shires, resulting in a total of 1088 ET-Minion cores. By software convention, 32 shires operate in parallel as a \emph{compute array}, one acts as a master Minion shire and one is held as a spare, although all 34 are architecturally identical. The compute array therefore provides $32 \times 32 \times 2 = 2048$ harts to kernels. Eight additional memory shires, four per side, carry sixteen 16-bit LPDDR4X physical interfaces paired into eight 32-bit controllers, operating at up to 4266~MT/s for an aggregate 133~GiB/s and up to 32~GiB of external memory. A PCIe Gen4 shire and an I/O shire which contains the ET-Maxions, the service processor and the root of trust complete the device. All shires are joined by a mesh Network-on-Chip (NoC) organised as an $8 \times 6$ grid with 44 occupied mesh stops, the corners being unpopulated. The total on-die SRAM is 140~MiB, which as mentioned is distributed across the shires.
The cache hierarchy is not coherent between levels. Standard RISC-V loads and stores operate at the per-Minion L1D level and Esperanto specific \texttt{l} and \texttt{g} instruction suffixes direct an access instead to operate at the shire cache or at the chip-level distributed L3 level, bypassing the levels beneath~\cite{etsoc1_prm}. Consequently, software is therefore responsible for placing each memory access at the level where the data is expected to reside, and for explicit eviction where data must become visible to other Minion(s).

\subsubsection{Theoretical Performance Bounds}

Peak throughput scales with Minion count and not with the 2048 harts the array exposes. At the nominal 600\,MHz operating point this gives $8 \times 2 \times 600\,\mathrm{MHz} = 9.6$\,GFLOP/s per core and, over the 32 compute
shires of the standard software configuration \cite{esperanto_etsoc1_datasheet, ditzel2022hardware}, an array peak of $P_{\mathrm{fp32}} = 9830$\,GFLOP/s.  This figure can fall to around 5243~GFLOP/s in certain kernel designs as described in section \ref{sec:simd}.

\subsubsection{A Packed-Single Vector ISA}

Each ET-Minion implements Esperanto's bespoke PS instruction set. In this extension the floating-point registers are natively 256 bits wide, and whereas standard RISC-V \texttt{F} instructions ignore the high 224 bits on input and zero them on
output, by contrast PS instructions operate lane-wise on all eight 32-bit lanes~\cite{etsoc1_prm,esperanto_etsoc1_datasheet}. The two harts of a Minion have separately banked register files but share the single eight-lane execution datapath.

The most applicable instructions from this PS ISA extension are:

\begin{itemize}
\item \texttt{fbc.ps fd, 0(rs)} loads a single fp32 scalar value and broadcasts it to all eight lanes of \texttt{fd}.
  \item \texttt{flw.ps fd, offset(rs)} loads 256 bits (eight fp32 values) from memory into \texttt{fd}.
  \item \texttt{fsw.ps fs, offset(rs)} stores 256 bits to memory from \texttt{fd}.
  \item \texttt{fmadd.ps fd, fs1, fs2, fd} undertakes an eight-lane fused multiply-add, $\text{fd} \leftarrow \text{fs1} \times \text{fs2} + \text{fd}$.
  \item \texttt{fxor.pi fd, fd, fd} writes zero to a PS register, effectively undertaking bitwise XOR with itself.
  \item \texttt{mov.m.x m0, zero, 0xff} sets the lane-enable mask register to enable all eight lanes, there are eight 8-bit mask registers, \texttt{m0}--\texttt{m7}.
\end{itemize}
Assembly code is added to the source using GCC inline assembly, and it should be noted that the ET GCC toolchain does not auto-vectorise using ET-SoC-1 PS instructions.

There are two deviations from the IEEE~754 floating-point arithmetic standard by the ET-SoC-1 which are relevant to numerical comparison. Denormal operands are treated as sign-preserved zero on input and flushed to sign-preserved zero on output before rounding, and floating-point division and square root are not implemented in hardware~\cite{etsoc1_prm}. Neither
of these affect the sGEMM inner loop directly, but both do impact the accuracy of some element-wise operations when comparing to a host reference code.

\subsubsection*{TensorFMA32 unit}

At the time of writing there are currently several active matrix extension standardisation efforts going on in RISC-V International by the community. The two most mature, which stand in contrast to each other, are Attached Matrix Extension (AME) and Integrated Matrix Extension (IME). AME assumes a separate matrix unit, with its own dedicated registers, datapath, and compute functionality. IME assumes that the matrix unit will be part of an existing RISC-V Vector (RVV) unit, sharing registers and datapath with that unit. By contrast the ET-SoC-1 takes a slightly different approach, where each ET-Minion contains a tensor micro-operation sequencer rather than a separate matrix-multiply unit. For fp32 operands this sequencer drives the same eight-lane FMA circuitry used by the PS SIMD instructions, issuing eight scalar multiply-accumulate operations per cycle, a $C_{16 \times 16} \mathrel{+}= A_{16 \times K_c} \times B_{K_c \times 16}$ product with $K_c = 16$ comprises 4096 scalar products and therefore occupies the datapath for 512 cycles~\cite{etsoc1_prm}. Half-precision and 8-bit integer variants (\texttt{TensorFMA16A32}, \texttt{TensorIMA8A32}) achieve two and eight times this rate respectively, but are not used in this work.

It is important to highlight that these tensor operations are not RISC-V instructions. Instead, they are issued by writing 64-bit command words to dedicated Control and Status Registers (CSRs) with \texttt{csrrw}, with some forms drawing a further 64 bits from \texttt{x31} and so behaving as effectively 128-bit instructions~\cite{etsoc1_prm}. Because only hart~0 within each Minion core may write these CSRs, only hart~0 can issue tensor operations and attempts from hart~1 will raise an illegal-instruction exception. The sole exception is \texttt{TensorLoadL2Scp}, where hart~1 may issue and which copies data from memory into shire scratchpad, acting as an explicit prefetch.

When performing matrix multiplication the \texttt{A} operand is read from the 3~KiB tensor register file exposed when the L1 data cache is placed in scratchpad mode, addressed as \texttt{L1Scp[0]}--\texttt{L1Scp[47]} with each entry holding up to 64 bytes. The \texttt{B} operand may be read either from \texttt{L1Scp} or from \texttt{TenB}. \texttt{TenB} is not a permanent architectural state but instead the \texttt{TensorLoadB} instruction forwards loaded data to the next operation that reads \texttt{TenB}, rather than committing it to a register file~\cite{etsoc1_prm}. For fp32 results, \texttt{C} occupies the floating-point registers of hart~0, which are a pair of registers holding each 16-element row, so that \texttt{f0}--\texttt{f31} hold the full $16 \times 16$ tile. Integer accumulation is different and instead uses the separate \texttt{TenC} registers and is not exercised in this paper.

Tensor micro-operations are emitted in program order but execute asynchronously with respect to the issuing thread, effectively behaving as if it were a third hardware thread. The sequencer does not track all hazards, so explicit wait operations are required to be provided by the programmer, both to observe completion and to separate certain conflicting pairs of tensor operations~\cite{etsoc1_prm}.

\subsubsection*{Block multithreading}

The ET-SoC-1 implements \emph{coarse-grained block multithreading} where when a hart stalls on a memory miss, the hardware switches execution to the other hart in the same Minion. The switch overhead is understood to be zero, with the hardware switching on the first stall cycle. This bears some resemblance to the approach adopted by GPUs, where Streaming Multiprocessors (SMs) switch between warps on stall. The ET-SoC-1 approach isn't so flexible, as there are only two harts per Minion core, so if both are stalled on a memory access then the core must wait, but this does potentially still help hide the overhead of data access.

The 4~KiB L1 data cache in each Minion core has three configurations~\cite{etsoc1_prm}:

\begin{enumerate}
    \item \textbf{Shared} where both harts share all 4~KiB
    \item \textbf{Split} in which hart~0 receives 2~KiB and hart~1 receives 512~bytes.
    \item \textbf{Scratchpad} each hart retains a 512-byte data cache and the remaining 3~KiB is allocated to the tensor register file.
\end{enumerate}

The SIMD kernel described in this paper runs in shared mode, giving one hart a natural mechanism to pre-populate the cache for the other. The tensor kernel that we have developed and runs in scratchpad mode.

\subsubsection{Single-Precision GEMM}

The Level-3 BLAS routine \texttt{sGEMM} computes $C \leftarrow \alpha A B + \beta C$ for row-major fp32 matrices $A \in \mathbb{R}^{M \times K}$, $B \in \mathbb{R}^{K \times N}$, and $C \in \mathbb{R}^{M \times N}$ with scalar parameters $\alpha, \beta \in \mathbb{R}$~\cite{dongarra1990level3}, extending the earlier Level-1 and Level-2 sets~\cite{lawson1979basic,dongarra1988blas}. The dominant cost is the inner product loop, requiring $2MNK$ floating-point operations.

High-performance GEMM implementations rely on register blocking, a $\mathrm{MR} \times \mathrm{NR}$ tile of $C$ is held in floating-point registers throughout the $K$ inner loop, avoiding repeated loads and stores to the $C$ matrix~\cite{goto2008anatomy,lam1991cache}. For the PS ISA extension, where each register holds eight fp32 values, a $4 \times 32$ block (MR$=4$ rows, NR$=32$ columns) occupies exactly the 32 available floating-point registers which comprises 16 accumulators, 4 vectors for operand \emph{B}, and 8 broadcasts for operand \emph{A} plus the next step's operands in a two-step pipeline.


\section{Device Kernels}

\subsection{Packed-Single SIMD Baseline}
\label{sec:simd}
The sGEMM code we developed is a $4 \times 32$ register-blocked microkernel (MR $= 4$, NR $= 32$) with inline assembly using the ET-specific PS ISA extension. The tile $C[i_0{:}i_0{+}4,\; j{:}j{+}32]$ accumulates into f0--f15, where $f[4r{+}v]$ holds row $r$ of PS vector $v$, giving $16 \times 8 = 128$ fp32 accumulators. The inner $K$ loop is \emph{two-step software pipelined}~\cite{lam1988software}, which claims the remaining 16 registers as two sets of operands and so consumes the entire FP register file, as shown in Algorithm~\ref{alg:simd}.

\begin{algorithm}
\caption{PS SIMD $4 \times 32$ microkernel, two-step software
pipelined. Each numbered line is fully unrolled straight-line code, so
line order is issue order. The loop body totals 59 instructions plus
the back-edge branch.}
\label{alg:simd}
\begin{algorithmic}[1]
\State $m_0 \gets \mathtt{0xff}$
       \Comment{\textbf{\textit{\texttt{mov.m.x}, enable all 8 lanes}}}
\State $c_{r,v} \gets 0,\ (r,v) \in [0,4)^2$
       \Comment{\textbf{\textit{16 $\times$ \texttt{fxor.pi}, f0--f15}}}
\For{$k \gets 0$ \textbf{to} $K-1$ \textbf{step} $2$}
  \State $\hat{a}^{0}_{r} \gets A[i_0{+}r,\,k],\ r \in [0,4)$
         \Comment{\textbf{\textit{4 $\times$ \texttt{fbc.ps} $\rightarrow$ f20--f23}}}
  \State $b^{0}_{v} \gets B[k,\ j{+}8v\,{:}\,j{+}8v{+}8],\ v \in [0,4)$
         \Comment{\textbf{\textit{4 $\times$ \texttt{flw.ps} $\rightarrow$ f16--f19}}}
  \State advance $A$, $B$ to $k{+}1$
         \Comment{\textbf{\textit{5 integer ops}}}
  \State $\hat{a}^{1}_{r} \gets A[i_0{+}r,\,k{+}1],\ r \in [0,4)$
         \Comment{\textbf{\textit{4 $\times$ \texttt{fbc.ps} $\rightarrow$ f24--f27}}}
  \State $b^{1}_{v} \gets B[k{+}1,\ j{+}8v\,{:}\,j{+}8v{+}8],\ v \in [0,4)$
         \Comment{\textbf{\textit{4 $\times$ \texttt{flw.ps} $\rightarrow$ f28--f31}}}
  \State advance $A$, $B$ to $k{+}2$; decrement trip counter
         \Comment{\textbf{\textit{6 integer ops}}}
  \State $c_{r,v} \mathrel{+}= \hat{a}^{0}_{r} \cdot b^{0}_{v},\
         (r,v) \in [0,4)^2$
         \Comment{\textbf{\textit{16 $\times$ \texttt{fmadd.ps}, step $k$}}}
  \State $c_{r,v} \mathrel{+}= \hat{a}^{1}_{r} \cdot b^{1}_{v},\
         (r,v) \in [0,4)^2$
         \Comment{\textbf{\textit{16 $\times$ \texttt{fmadd.ps}, step $k{+}1$}}}
\EndFor
\State store $c_{r,v}$ to \textit{tile}
       \Comment{\textbf{\textit{16 $\times$ \texttt{fsw.ps}}}}
\end{algorithmic}
\end{algorithm}

Software pipelining is effective on this core because, although the Minion is single-issue and in-order, a load blocks at the first instruction that consumes its result rather than at the load itself. Separating a load from its consumer therefore converts load latency into useful issue slots. Lines 7--8 issue the step-$k{+}1$ operands early enough to remain in flight throughout the 16 FMAs of line 10.

The instruction mix also imposes a ceiling independent of any memory stall. Summing Algorithm~\ref{alg:simd} gives 60 instructions per iteration, of which only 32 are FMAs, so a stall-free execution would reach $32/60 \approx 53\%$ of arithmetic peak, or approximately 5243~GFLOP/s. The measured 777~GFLOP/s is 14.8\% of that issue-limited ceiling, confirming that the shortfall is dominated by stalls rather than by instruction mix. The $\alpha/\beta$ epilogue, $C[i,j] \leftarrow \alpha \cdot \mathrm{tile}[i,j] + \beta \cdot C[i,j]$, is applied in scalar C.

\subsection{TensorFMA32 Kernel}

The tensor kernel uses a $16 \times 16$ output tile (MT$=16$, NT$=16$), matching the  native TensorFMA32 tile geometry. The $K$ dimension is consumed in chunks of $K_c = 16$ and $C$ accumulates in the vector register file throughout.  Algorithm~\ref{alg:tensor} gives the resulting chunk loop.

\begin{algorithm}
\caption{TensorFMA32 microkernel: $16 \times 16$ output tile with
double-buffered $A$. $K_c = 16$, and buffer $b$ occupies
\texttt{L1Scp}$[16b \ldots 16b{+}15]$.}
\label{alg:tensor}
\begin{algorithmic}[1]
\Require L1D in scratchpad mode; hart~0 of the Minion; $K \bmod K_c = 0$
\State \texttt{TensorLoad} $A_0 \to$ buffer $0$; \ \texttt{Wait}(load)
\For{$k = 0, \ldots, K/K_c - 1$}
  \If{$k + 1 < K/K_c$}
    \State \texttt{TensorLoad} $A_{k+1} \to$ buffer $(k{+}1) \bmod 2$
           \Comment{\textbf{\textit{overlaps the FMA below}}}
  \EndIf
  \State \texttt{TensorLoadB} $B_k \to$ \texttt{TenB}
         \Comment{\textbf{\textit{forward-paired, no wait}}}
  \State \texttt{TensorFMA32} $C \mathrel{+}= A_k B_k$; \ $A_k$ from
         buffer $k \bmod 2$, $B_k$ from \texttt{TenB}, \
         \textit{first\_pass} $= (k = 0)$
  \State \texttt{Wait}(fma)
         \Comment{\textbf{\textit{drains \texttt{f0}--\texttt{f31}, frees buffer $k \bmod 2$}}}
  \If{$k + 1 < K/K_c$}
    \State \texttt{Wait}(load) \Comment{\textbf{\textit{$A_{k+1}$ now resident}}}
  \EndIf
\EndFor
\State spill $C$ from \texttt{f0}--\texttt{f31}
\end{algorithmic}
\end{algorithm}

 Scratchpad mode is entered by setting bits $[1:0]$ of CSR \texttt{0x810} (\texttt{UCACHE\_CONTROL}) to \texttt{0b11}, asserting \texttt{d1\_split} and \texttt{scp\_en}~\cite{etsoc1_sdk}, preceded by a \texttt{fence} and a \texttt{WAIT\_CACHEOPS} on CSR \texttt{0x830}. This exposes 3~KiB as \texttt{L1Scp[0]}--\texttt{L1Scp[47]} and reduces each hart's data cache to 512~bytes; the transition zeroes the scratchpad, which is benign since every tile reloads $A$. Double-buffering $A$ consumes 32 of the 48 entries, so a third buffer would fit, though it is not exploited here. Double-buffering $B$ as well would need 64 entries, obtainable only by shrinking the tile and with it the arithmetic intensity, so $B$ streams through \texttt{TenB} instead. The cost is that \texttt{TensorLoadB} issues immediately before the FMA consuming it, leaving the $B$ gather with no prefetch distance, which Section~\ref{sec:tensor-eff} identifies as the dominant cost in the measured result. Only hart~0 of each Minion may issue tensor operations, so the kernel runs on 1024 harts rather than 2048, with hart~1 idle throughout.


\section{Benchmark Methodology}

\subsection{Problem Specification}

All benchmarks compute $C = \alpha A B + \beta C$ with $\alpha = 1.0$, $\beta = 0.5$,  and $M = N = K$ chosen from $\{2048, 4096\}$. The $2048^{3}$ problem ($2 \times 2048^3 \approx 17.2~\text{GFLOP}$) was selected because its aggregate working set, 48~MiB for three $2048 \times 2048$ fp32 matrices, fits within the 136~MiB of shire cache available across the compute array. Consequently, a well-distributed kernel can hold most of the data on chip. The $4096^{3}$ problem, which is approximately 137~GFLOP, 192~MiB working set, deliberately exceeds on-chip capacity, and thus stressing DRAM bandwidth. Both sizes are multiples of 32,  satisfying the kernel's alignment requirement.

\subsection{Parallel Tile Distribution}

The two device kernels that we focus on in this paper distribute work differently, because they have different operand-reuse structure and different numbers of eligible harts. 

The SIMD kernel uses all 64 harts of each shire, each iterating a tile with a stride of 64. The stripe width is chosen so that the shire's $B$ working set stays cache-resident.

The tensor kernel can only use hart~0 of each Minion, so this kernel has 1024 eligible harts rather than 2048. It assigns tiles from a single flat chip-wide index over those harts, striding by 1024, rather than from a per-shire stripe. Consecutive tiles are consecutive column tiles of one row-block, so the 32 Minions of a shire hold 32 consecutive column tiles of the same row-block and share its $A$ panel 32 ways through their common shire cache. Measurement showed this $A$ reuse is dominant so restricting each shire to an L2-resident $B$ stripe, as the SIMD kernel does, lost more on $A$ than it gained on $B$. 

\subsection{Shire B-Striping}

The B matrix ($K \times N$) cannot fit in any single shire's 4~MiB L2 for  $N \geq 2048$ ($K \times N \times 4 = 16~\text{MiB}$ per shire if assigned  the full matrix). Instead, each shire $s$ is assigned a contiguous stripe  of $\lceil N_T / N_S \rceil$ column tiles, where $N_T = N / \mathrm{NR}$ is the total column-tile count and $N_S = 32$ is the shire count. To fall within the 4~MiB shire L2 cache, the B working set per shire is restricted to:

\begin{equation*}
\begin{split}
S_B &= K \times \left\lceil\frac{N}{N_S \cdot \mathrm{NR}}\right\rceil
       \times \mathrm{NR} \times 4 \\
    &\approx 512~\text{KiB} \quad (N = 2048), \\
    &\approx 2~\text{MiB} \quad\;\, (N = 4096).
\end{split}
\end{equation*}

\subsection{Reference Implementations}

We compare performance on the ET-SoC-1 against five reference implementations on other architectures.

\subsubsection{Hand-written CPU kernel}
This is a register-blocked microkernel (MR$=8$, NR$=24$) vectorised with RVV 1.0 vector intrinsics on SpacemiT K3, AVX-512 FMA on  x86-64, or an NR$=12$ variant using 128-bit fixed-length SVE2~\cite{stephens2017arm} on aarch64. The  accumulator tile is held in ZMM (512-bit) or SVE registers across the entire $K$  inner loop, eliminating repeated C traffic. OpenMP is used for parallelisation  across physical cores.

\subsubsection{OpenBLAS}
The primary library reference~\cite{zhang2012openblas,openblas_repo}.  Its GEMM implementation uses hand-tuned assembly microkernels and  BLIS-style~\cite{vanzee2015blis} packing loops to exploit the full memory hierarchy.

\subsubsection{NVIDIA cuBLAS}
On Nvidia GPUs, with the CUDA math library providing GEMM for our runs on an NVIDIA GB10 Blackwell GPU (the DGX Spark), and in this paper evaluated in both fp32 and TF32 precision.

\subsubsection{AMD hipBLAS}
For AMD GPUs, the ROCm math library provides GEMM and we leverage this on an AMD Radeon RX~9060~XT (RDNA~4~\cite{amd2025rdna4}, gfx1200 per ROCm 7.1.0), evaluated in fp32 mode.

\subsection{Numerical Verification}

In this work OpenBLAS serves as the primary verification reference where each device's result is verified 
element-wise against the OpenBLAS output with a tolerance:
\[
|d_i - r_i| \leq \mathrm{atol} + \mathrm{rtol} \cdot |r_i|,
\]
where $\mathrm{atol} = 10^{-4} + 2 \times 10^{-7} K$ and $\mathrm{rtol} = 10^{-3}$. The absolute tolerance scales with $K$ because fp32 accumulation error grows as  $K \cdot \varepsilon_{\mathrm{mach}}$ for near-cancellation elements, a fixed  $\mathrm{atol}$ of $10^{-4}$ lies below the fp32 noise floor at $K = 4096$,  where even OpenBLAS and cuBLAS disagree on a single element~\cite{higham2002accuracy}.

A further contribution on the ET-SoC-1 is the flushing of denormal operands and results to sign-preserved zero, which precludes bit-exact agreement with a host reference in the presence of near-cancellation.

GPU variants operating in TF32 or FP16 precision use a looser tolerance ($\mathrm{atol} = 0.1$, $\mathrm{rtol} = 0.02$). These formats round the multiplicands to approximately 10-bit mantissas before the multiply, even where accumulation is performed in fp32, and so are not expected to agree with the fp32 reference to the same tolerance as the device kernels.


\section{Experimental Results}
\subsection{Experimental Platform}

All ET-SoC-1 measurements were collected on the AIFoundry community server, hosting a PCIe-attached ET-SoC-1 card. The device operates at 600~MHz with all 32 compute shires active (shire mask 0xffffffff), providing 2048 active harts. The reported \emph{compute only} time excludes PCIe data transfers between the host and device and the \emph{total} time includes both on-card compute and all data transfers. GFLOP/s is derived as $2MNK / t$, where $t$ is the host-side wall-clock elapsed time measured with \texttt{std::chrono::steady\_clock} bracketing the kernel launch and its completion synchronisation barrier (\texttt{waitForStream}).

The hardware characteristics are summarised in Table\ref{tab:platforms}.  All code was written in GCC 15 and built with CMAKE.  Libraries used were OpenBLAS 0.3.32, CUDA 13.0 and ROCm 7.1.  OpenBLAS and OpenMP use maximum avalibale CPU threads. 

\begin{table}[ht]
\centering
\caption{Evaluated platforms. Threads are those used by OpenMP and OpenBLAS.}
\label{tab:platforms}
\footnotesize
\setlength{\tabcolsep}{3pt}
\begin{tabular}{lrrlrr}
\hline
Platform & Cores (thr.) & Clock & SIMD \\
\hline
ET-SoC-1          & 1024 (2048) & 0.6~GHz  & 256-bit PS  \\
Intel i5-11600    & 6 (12)      & 2.8~GHz  & AVX-512     \\
AMD Ryzen 9 9950X & 16 (32)     & 4.3~GHz  & AVX-512     \\
DGX Spark CPU     & 10+10 (20)  & 4.0~GHz  & SVE2 128-bit  \\
SpacemiT K3 X100  & 8 (8)    & 2.4~GHz  & RVV 1.0 256-bit   \\
SpacemiT K3 A100  & 8 (8)    & 2.4~GHz  & RVV 1.0 1024-bit  \\
\hline
NVIDIA GB10 GPU   & 48 SMs  & 2.4~GHz  & --           \\
AMD RX 9060 XT    & 32 CUs  & 2.5~GHz  & --          \\
\hline
\end{tabular}
\end{table}

The theoretical single-precision peak follows from the per-core datapath which is 1024 Minions $\times$ 8 lanes $\times$ 2 FLOP per FMA $\times$ 600~MHz $= 9830$~GFLOP/s. Both device kernels are bounded by this same figure, since the tensor sequencer drives the same FMA lanes. The 9830~GFLOP/s figure assumes every issue slot carries an FMA, which the microkernel's instruction mix precludes. Because the ET-Minion is single-issue and in-order it retires at most one instruction per cycle and cannot co-issue integer work alongside floating-point, so every instruction that is not an \texttt{fmadd.ps} displaces one that could have been. A perfectly stall-free execution therefore spends only 32 of every 60 cycles on arithmetic, delivering $32 \times 8 \times 2 = 512$~FLOP per 60 cycles per Minion and hence $32/60 \approx 53.3\%$ of peak, or 5243~GFLOP/s across the 1024-Minion compute array.

\subsection{ET-SoC-1 Kernel Performance}

Table~\ref{tab:ETsoc1-perfomance} presents measured performance for the two ET-SoC-1 kernels at both problem sizes. It can be observed that the TensorFMA32 kernel attains 2676~GFLOP/s at $2048^{3}$, 3.44$\times$ the SIMD baseline of 777~GFLOP/s. At the larger $4096^{3}$ problem, both kernels degrade as the working set spills to LPDDR4x where TensorFMA32 drops to 1849~GFLOP/s and the SIMD baseline to 546~GFLOP/s.

\begin{table}[ht]
\centering
\caption{ET-SoC-1 compute-only performance (GFLOP/s)\\at 600~MHz, 32 shires, fp32.}
\label{tab:ETsoc1-perfomance}
\begin{tabular}{lrr}
\hline
Kernel & $N=2048$ & $N=4096$ \\
\hline
SIMD baseline & 777  & 546  \\
TensorFMA32   & 2676 & 1849 \\
\hline
\end{tabular}
\end{table}

\subsection{Cross-Platform Comparison}

Table~\ref{tab:crossplatform-performance} presents a full performance comparison across all evaluated platforms and implementations. For the smaller problem size the ET-SoC-1 SIMD baseline (777~GFLOP/s) outperforms all CPUs configurations apart from the AMD 9950X with OpenBLAS, and the TensorFMA32 result (2676~GFLOP/s) then closes the gap and outperforms this AMD 9950X OpenBLAS figure (2300~GFLOP/s). The larger problem size presents a more complex performance picture on the CPU, where the SIMD version is now outperformed by four out of nine configurations and the TensorFMA32 outperformed by the AMD 9950X with OpenBLAS.  Cache size plays a big part in performance seen in the hand-written register-blocked kernel at $N=4096$ problem size, the i5 (12MiB L3) and GB10 (8MiB L3) show a slowdown in performance but the 9950X (64 MiB L3) shows a speed up. 

When comparing against the GPU, the GB10 Blackwell GPU achieves 16\,400~GFLOP/s with cuBLAS in fp32 at $2048^{3}$, which is approximately 6.1$\times$ the ET-SoC-1 TensorFMA32's result. The RX~9060~XT with hipBLAS attains 4\,985~GFLOP/s, approximately 1.9$\times$ the TensorFMA32 result. Hand-written CUDA and HIP kernels are slower than their respective library references by a factor of roughly 2$\times$ (fp32) and 4$\times$ (tensor/WMMA). This is qualitatively consistent with long-standing reports of reference-versus-library gaps in the GPU GEMM literature~\cite{nath2010improved} with the vendors having invested significant effort to optimise libraries, although it is noted that no directly comparable recent study is available for these architectures. The TF32 and FP16 tensor variants are noted separately because they use reduced-precision arithmetic ($\approx$10-bit mantissa) and cannot be compared directly with the fp32 ET-SoC-1 results on a numerical basis~\cite{markidis2018nvidia}. However, ultimately all GPU configurations out perform both SIMD and TensorFMA32 on the ET-SoC-1 for both problem sizes.

\subsection{Energy Efficiency}

The final column of Table~\ref{tab:crossplatform-performance} reports energy efficiency in GFLOP/Watt at the $N = 4096$ problem size. These values use measured power rather than TDP.  The ET-SoC-1 card power was sampled with \texttt{et-powertop} during a sustained run of 100 back-to-back kernel launches, CPU package power was read from the Linux RAPL energy counters with idle power subtracted, and GPU board power was sampled with the vendor tools (\texttt{nvidia-smi} and \texttt{amd-smi}), with the exception of the DGX Spark CPU entries (marked \dag), which are calculated from documented power figures. The ET-SoC-1 has a much lower power draw than the other CPUs and GPUs within this study.  Against the CPUs the efficiency results are decisive.  The TensorFMA32 kernel 147.0~GFLOP/Watt which is around $7\times$ the best CPU result (the AMD 9950x with OpenBLAS at 20.6~GFLOP/Watt). The SIMD baseline reaches 68.3~GFLOP/Watt, or $3.3\times$ that figure, despite running at only 7.9\% of arithmetic peak.

\begin{figure*}[ht]
\centering
\includegraphics{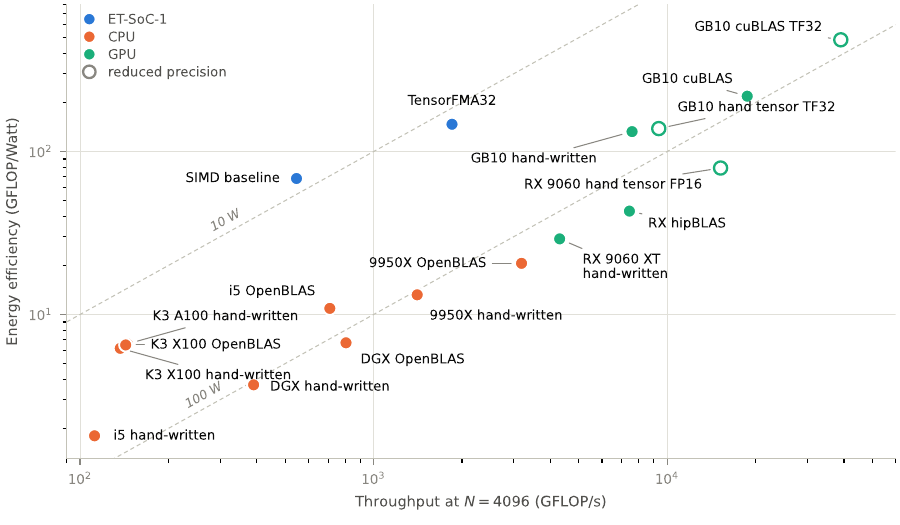}
\caption{sGEMM throughput against energy efficiency at $N = 4096$. Both
axes are logarithmic, and the diagonals are lines of constant power,
since GFLOP/Watt is throughput divided by power. Open markers denote
reduced-precision (TF32/FP16) results, which are not numerically
comparable with the fp32 figures. Values are those of Table~\ref{tab:crossplatform-performance}.}
\label{fig:efficiency}
\end{figure*}

Against the GPUs the outcomes are mixed depending on the comparisons being made.  Comparing like-for-like against hand-written kernels, the ET-SoC-1 and the NVIDIA GB10 GPU give GFLOP/Watt results that are within 10\% of each other. This changes when the cuBLAS kernel is used, the advantage of the NVIDIA produced library gives the GB10 GPU a distinct advantage which highlights a potential software tuning gap over an architectural limit would require a tuned BLAS library for the ET-SoC-1, which does not presently exist.  The AMD RX 9060 XT (a lower end gaming GPU) gives a much poorer result when compared to either the GB10 or the ET-SoC-1. See figure \ref{fig:efficiency}.

\begin{table*}[ht]
\centering
\caption{sGEMM compute-only performance (GFLOP/s) across platforms. GPU TF32/FP16 variants use reduced-precision \\ arithmetic and are marked accordingly. ``Hand-written'' entries are bespoke register-blocked microkernels, \\ ``library'' entries invoke tuned-library implementations. \\ GFLOP/Watt measured for the $N=4096$ problem size.}
\label{tab:crossplatform-performance}
\begin{tabular}{llrrr}
\hline
Platform & Implementation & $N=2048$ & $N=4096$ & GFLOP/Watt \\
\hline
ET-SoC-1 (600~MHz, 32 shires) & SIMD packed-single (baseline)    & 777   & 546   & 68.3 \\
ET-SoC-1 (600~MHz, 32 shires) & TensorFMA32                      & 2676  & 1849  & 147.0 \\
\hline
Intel i5-11600 (AVX-512)      & Hand-written register-blocked    & 228   & 112   & 1.8 \\
Intel i5-11600                & OpenBLAS                         & 596   & 709   & 10.9 \\
AMD Ryzen 9 9950X (AVX-512)   & Hand-written register-blocked    & 560   & 1407  & 13.2 \\
AMD Ryzen 9 9950X             & OpenBLAS                         & 2300  & 3190  & 20.6 \\
NVIDIA DGX Spark CPU (SVE2)   & Hand-written register-blocked    & 450   & 390   & 3.7\dag \\
NVIDIA DGX Spark CPU          & OpenBLAS                         & 740   & 805   & 6.7\dag \\
SpacemiT K3 X100 CPU (RVV)    & Hand-written register-blocked    & 72    & 137   & 6.2 \\
SpacemiT K3 A100 CPU (RVV)    & Hand-written register-blocked    & 145   & 145   & 6.6 \\
SpacemiT K3 X100 CPU          & OpenBLAS                         & 140   & 143   & 6.5 \\
\hline
NVIDIA GB10 Blackwell GPU     & Hand-written packed-single (fp32) & 7550  & 7590  & 132.5 \\
NVIDIA GB10 Blackwell GPU     & Hand-written tensor (TF32)*      & 9230  & 9360  & 138.3 \\
NVIDIA GB10 Blackwell GPU     & cuBLAS (fp32)                    & 16400 & 18730 & 218.6 \\
NVIDIA GB10 Blackwell GPU     & cuBLAS (TF32)*                   & 31800 & 39030 & 484.9 \\
AMD RX 9060 XT (RDNA 4)       & Hand-written packed-single (fp32) & 2926  & 4302  & 29.1 \\
AMD RX 9060 XT (RDNA 4)       & Hand-written tensor (FP16)*      & 10810 & 15200 & 79.2 \\
AMD RX 9060 XT (RDNA 4)       & hipBLAS (fp32)                   & 4985  & 7435  & 43.1 \\
\hline
\multicolumn{5}{l}{\footnotesize * Reduced precision ($\approx$10-bit mantissa) not numerically comparable with fp32 results. \dag Calculated from documented values}
\end{tabular}
\end{table*}


\section{Discussion}

\subsection{Dominant Bottleneck: A-Matrix DRAM Latency}

The SIMD baseline reaches 7.9\% of the 9830~GFLOP/s arithmetic peak, and 14.8\% of the 5243~GFLOP/s issue-limited ceiling derived in Section \ref{sec:simd}. The additional overhead is stall time, and its distribution between the two operands is not obvious from their working-set sizes alone.

Aggregate DRAM traffic is dominated by operand $A$. With shire-local $B$ stripes, each shire's $B$ working set is cache-resident, so $B$ is fetched from DRAM approximately once per shire. $A$, by contrast, is read by every shire, giving roughly $32 \times 16~\text{MiB} = 512~\text{MiB}$ over the 22.1~ms runtime, or about 23~GiB/s. This is consistent with the demand load reported in Section \ref{sec:arch_obs} and confirms that $A$ is the off-chip traffic source.

Memory access \emph{frequency}, however, is dominated by $B$. For each $k$-step the kernel reads 128 contiguous bytes of $B$, which is two cache lines and the next $k$-step lies a full row away, $N \times 4$ bytes distant, so no line is reused. That is four line fetches per loop iteration. $A$, in contrast, advances four bytes per $k$-step, so a 64-byte line serves 16 steps and the four $A$ rows together generate only 0.5 line fetches per iteration. $B$ is therefore fetched about eight times more often than $A$.

Comparing frequency with cost, $A$ dominates the stall budget only if $L_{\text{DRAM}} > 8 L_{2}$, that is, only if an off-chip access costs more than eight times a shire-cache hit. For plausible values of $L_{2} \approx 40$ cycles this requires $L_{\text{DRAM}} > 320$ cycles. The more likely position is that $B$'s frequent shire-cache traffic and $A$'s infrequent off-chip traffic contribute comparably, with $B$ the larger contributor. This revises the natural first reading of the working-set figures, and it changes which optimisation to focus on. Reducing $B$ fetch frequency by widening $\mathrm{MR}$, or steering $B$ loads past L1 with the cache-targeted \texttt{flwl.ps} form so that streaming $B$ lines do not evict the $A$ working set from the 4~KiB L1, are both aimed more directly at the dominant cost than deepening the software pipeline.

It is noted that neither approach closes the gap fully. For instance, at $s = 22$ slots and 0.5 $A$ fetches per 60-instruction iteration, a purely latency-driven model predicts a substantially higher throughput than the 777~GFLOP/s that was measured, for any $L_{\text{DRAM}}$ in the plausible range. The additional overhead is likely to lie in outstanding-miss capacity, where the kernel issues eight loads back-to-back and assumes all are concurrently in flight or in contention between the two harts of a Minion for the single load-store path. Distinguishing these requires the per-neighbourhood performance counters and is left as future work.

\subsection{Deep Software Pipelining}

The analysis above raises the question of whether increasing pipeline depth, and so widening the window between a load's issue and its first consumer, could hide a greater fraction of memory latency. Let $L$ be the load-to-use latency of a DRAM-resident line in core cycles, and $s$ the number of instruction-issue slots between a load and its first consumer, taken as the minimum over the loads of one iteration. A pipeline with $s$ slots hides $\min(s/L,\,1)$ of each stall, leaving $L - s$ cycles exposed. A published figure for $L$ for the ET-SoC-1 was not found, and it may not exist for this device, since the datasheet's AC electrical characteristics are not yet populated and the ET-Minion implements neither the RISC-V \texttt{Zicntr} extension nor any other cycle counter. Direct measurement would require either a long dependent-load chain timed against the 10~MHz RV timer or the per-neighbourhood performance counters. The analysis is therefore expressed as a function of $L$.

With MR $= 4$ the 32 FP registers are exactly consumed, by 16 accumulators and $2 \times (4~\text{A broadcasts} + 4~\text{B vectors}) = 16$ operands. Reducing MR to 2 halves the accumulators to eight and leaves each of four pipeline steps six operand registers, again filling the file exactly. The window widens by less than that fourfold depth suggests, however, because halving MR also halves each step's FMA block from 16 to 8. Applying the definition above to the two schedules gives $s = 22$ for the two-step kernel, set by its step-$k$ $B$ vectors, against $s = 31$ for the four-step, set by the final step's $B$ vectors, a factor of 1.4, not 4.

A prototype was compiled and run on hardware, and produced incorrect results with 4\,194\,099 of 4\,194\,304 elements falling outside tolerance at $N = 2048$, with a maximum absolute error of 77.9 against the CPU reference. Its compute time of 1.3~ms implies 13\,041~GFLOP/s, which exceeds the 9830~GFLOP/s arithmetic peak and is therefore impossible for a correct computation, strongly suggesting that the kernel terminates the $K$ loop prematurely or otherwise computes a partial result. As yet, the defect has not been isolated.

Taken together these figures settle the question without measuring $L$. The exposed stall per miss falls only from $L - 22$ to $L - 31$, a bound of $1.13\times$ at $L = 100$ and $1.02\times$ at $L = 400$, which the $0.86$ reduction in the issue-limited ceiling more than cancels for any $L$ above approximately 85 core cycles, or 142~ns at 600~MHz.  This is below any plausible off-chip access traversing the mesh interconnect. That is before the doubled $B$ traffic is counted at all. The four-step pipeline is therefore a net loss at any credible latency and so the prototype defect is probably not worth pursuing on performance grounds, and prefetching $B$ into shire scratchpad, which attacks the traffic term rather than the latency term, would be a promising direction to investigate.

\subsection{TensorFMA32 Efficiency}
\label{sec:tensor-eff}

The TensorFMA32 kernel achieves 27\% of the 9.83~TFLOP/s theoretical peak. Because the tensor sequencer drives the same eight-lane FMA circuitry as the PS SIMD path, the $3.44\times$ improvement cannot be attributed to additional arithmetic capability as both kernels contend for an identical peak. The gain is therefore one of operand delivery and issue efficiency. A single tensor operation replaces 256 individually issued \texttt{fmadd.ps} instructions and their associated loads, so that 512 cycles of arithmetic proceed  from a single instruction issue, whilst \texttt{A} is served from the on-core scratchpad and \texttt{B} is forwarded directly to the multipliers rather than being staged through the general-purpose load-store path. 

The operands do not, however, simply remain in registers. Both kernels hold $C$ in the register file for the whole $K$ loop, f0--f15 for the SIMD tile and f0--f31 for the tensor tile, so $C$ traffic distinguishes neither. The tensor unit's $A$ operand resides in scratchpad rather than in registers. What improves is the operand cost per unit of arithmetic. A $\mathrm{MR} \times \mathrm{NR}$ tile consumes $\mathrm{MR} + \mathrm{NR}$ operand elements per $k$-step whilst performing $2\,\mathrm{MR}\,\mathrm{NR}$ FLOP, an intensity that is maximised when the tile is square: the SIMD kernel's $4 \times 32$ geometry yields 1.8 FLOP per byte against 4.0 for the tensor unit's $16 \times 16$. 

The squarer geometry rebalances reuse rather than raising it uniformly. Each $A$ element feeds 32 columns in the $4 \times 32$ tile but only 16 in the $16 \times 16$, whilst each $B$ element goes from four uses to sixteen, and it is equalising the two that raises the intensity. Operand $A$ is also delivered by a single \texttt{TensorLoad} per chunk rather than by an \texttt{fbc.ps} issue per row at every $k$-step, so the same data costs one instruction per sixteen $k$-steps instead of four per step.

The 73\% gap from that achieved to the arithmetic peak is not attributable to loop level overhead. Each chunk issues around 12 instructions which comprises 1 \texttt{TensorLoad}, 1 \texttt{TensorLoadB} setup, 1 \texttt{TensorFMA32}, 2 waits, and the
loop arithmetic, against 512 cycles of tensor execution, which accounts for approximately 3\%.

We found that the dominant cost is operand gathering where each chunk's $A$ panel comprises 16 cache lines strided by $K \times 4$ bytes, and each $B$ panel a further 16 lines strided by $N \times 4$ involving 32 strided line fetches per 512-cycle arithmetic phase. At the measured 27\% efficiency the loop occupies approximately 1880 cycles per chunk, of which some 1370 are not covered by the FMA, which is about 43 cycles per fetch. This is consistent with partially pipelined shire-cache access that a single overlapped FMA cannot fully hide.

Two structural properties of the present implementation follow directly. First, only $A$ is double-buffered, and \texttt{TensorLoadB} is issued immediately before the \texttt{TensorFMA32} that consumes it, with no prefetch distance, so the entire $B$ panel gather is exposed. Extending double-buffering to $B$ is the most direct available remedy. Secondly, the arithmetic intensity of the chunk loop is 8192~FLOP per 2~KiB, or 4~FLOP per byte. Sustaining 2676~GFLOP/s at that intensity would require 669~GiB/s which is five times the LPDDR4X peak. The tensor kernel at $N = 2048$ is therefore served predominantly from the shire and chip-level caches rather than from DRAM, which is consistent with the $A$-panel sharing that the flat tile distribution is designed to exploit. At $N = 4096$ the working set no longer permits that residency, and the 31\% reduction to 1849~GFLOP/s reflects the memory overhead when transitioning to off-chip data held in external memory.

A further inefficiency is that hart~1 of every Minion is idle throughout the tensor kernel, since only hart~0 may issue tensor operations. The one tensor operation available to hart~1, \texttt{TensorLoadL2Scp}, copies data from arbitrary memory into the shire's scratchpad and so functions as a software-directed prefetch. Given that operand gathering is the binding constraint, using the 1024 otherwise-idle harts to stage panels into a shire's scratchpad ahead of the tensor pipeline is the most direct optimisation the architecture affords, and is the principal avenue for future work.

\subsection{Architectural Observations}
\label{sec:arch_obs}
Several architectural characteristics of the ET-SoC-1 have been highlighted by the experimental results. Firstly, \emph{the PS SIMD unit is latency-bound, not bandwidth-bound} where the DRAM demand load at 20~GiB/s is only 15\% of the 133~GiB/s peak. The bottleneck is in the per-request latency combined with the inability of the 2-step pipeline  to overlap enough instruction work to hide it. Consequently, the binding constraint is the per-hart request rate, not aggregate memory bandwidth~\cite{williams2009roofline}.

\emph{Operand staging, not arithmetic width, separates performance of the two kernels.}
By holding $A$ in on-core scratchpad and forwarding $B$ directly to the multipliers, the tensor path removes most per-operand traffic from the general-purpose load-store unit and reduces instruction issue to one operation per 512 cycles of arithmetic. The SIMD kernel, by contrast, must issue 16 loads and 32 FMAs every 60-instruction iteration and sustain them through the general-purpose load-store path. Both drive the same eight-lane datapath against the same 9830~GFLOP/s peak, so the $3.44\times$ difference is entirely one of delivery. 

\emph{Two-way multithreading is insufficient to hide DRAM latency.} Swapping out warps that stall on memory access within an SM on GPUs is an effective technique, with modern GPUs able to interleave sixteen or more warps per
scheduler. However, with only two harts per Minion to round-robin between, the ET-SoC-1 has roughly an order of magnitude less thread-level latency-hiding capacity than a contemporary GPU's SM. Two threads can cover a stall only if the second finds enough independent work to fill it. Due to the latency associated with DRAM-resident memory access it often cannot, and the Minion core stalls.

\subsection{Feasibility of Production Library Techniques}
GEMM libraries included in products such as OpenBLAS and products based on the BLIS framework \cite{goto2008anatomy,vanzee2015blis} often improve on native register-blocked kernels primarily through data packing and multi-level cache blocking.  Both of these techniques are directly applicable to the TensorFMA32 kernel.  The bottleneck analysis in Section~\ref{sec:tensor-eff} shows the areas where each would act.

Data packing eliminates the strided memory access pattern that accounts for the dominant stall cost. In the present kernel, the $B$ panel is gathered with stride $N \times 4$ bytes per row, producing 32 non-consecutive cache-line fetches per 512-cycle arithmetic phase. Pre-packing $B$ into a contiguous shire-local buffer prior to the $K$ loop would convert these strided fetches to sequential loads, removing the gather penalty. The second hart (hart~1) of each core can each issue \texttt{TensorLoadL2Scp} instructions to copy data into the shire scratchpad. If this packing is overlapped with tensor execution on hart~0, the additional data operation would hopefully not increase wall-clock time.

Multi-level cache blocking is already partially realised by the shire-stripe $B$ distribution, which keeps each shire's $B$ working set shire-cache-resident at $N = 2048$. At $N = 4096$ the 192~MiB aggregate working set exceeds the 136~MiB of on-chip SRAM, and the measured 31\% throughput reduction reflects the resulting DRAM traffic. Explicit blocking for the chip-level SRAM hierarchy, which is comparable to the $k_c$, $m_c$, $n_c$ tile parameters of the BLIS framework, would bound the working set to on-chip capacity across both problem sizes, substantially reducing the DRAM communication overhead for large matrices. Together, these two optimisations represent the most structurally targeted path towards closing the gap between the present kernel and a tuned library implementation.

\section{Conclusions}
\label{sec:conclusion}

This paper has presented a performance characterisation of the ET-SoC-1
using two sGEMM device kernels. The tensor micro-operation sequencer provides 3.4$\times$ the throughput of the SIMD baseline (2676 vs.\ 777~GFLOP/s at $2048^{3}$), achieving 27\% of the 9.83~TFLOP/s theoretical peak. Since both kernels drive the same eight-lane FMA datapath, this gain is attributable wholly to operand staging and issue amortisation rather than to additional arithmetic capability.

The SIMD baseline is memory-latency-bound rather than memory-bandwidth-bound. 2048 simultaneously stalling harts generate only $\approx$20~GiB/s of DRAM traffic (15\% of the 133~GiB/s nominal peak), confirming that the binding constraint is per-request latency, not aggregate bandwidth. A 4-step pipeline with MR$=2$ is theoretically feasible within the 32-register budget but this would potentially only increase the latency-hiding window to $1.4\times$ which does not pay for the doubled $B$ traffic. The prototype implementation produced incorrect results, diagnosis remains future work.

The ET-SoC-1 occupies an unusual position relative to contemporary hardware as it provides competitive SIMD throughput per watt at low area cost, but cannot match the absolute GEMM throughput of modern GPU tensor units or tuned CPU libraries at $N > 2048$ due to the DRAM latency bottleneck in its in-order scalar core design. The TensorFMA32 unit brings its performance into a competitive range for the smaller problem size, where its true fp32 accuracy provides a precision advantage over GPU TF32 and FP16 tensor implementations.

There are several possible directions that could follow on from this work.  The most obvious step to continue developing the code written for this project would be the introduction of per-neighbourhood performance monitoring. Measuring load-to-use latency and stall-cycle counts directly could help shape future optimisation.

In the tensor kernel, hart~1 of every Minion is idle, and the one tensor operation available to it,  \texttt{TensorLoadL2Scp}, copies data from memory into shire scratchpad and so functions as a software-directed prefetch therefore using the 1024 otherwise-idle harts to stage panels ahead of the tensor pipeline addresses the  gathering cost identified above.

Exploitation of the unused third of the \texttt{L1Scp} register file to triple-buffer \texttt{A} could well improve performance.  Further diagnosis and correction of the 4-step pipeline design might give improved results, but a potential of $1.4\times$ rather than doubling is assumed at this point, which may not be realised when set against the the doubled $B$ traffic. 

Beyond the present device, the comparison would benefit from extending the tensor kernel to batched GEMM and to the sequencer's native fp16 and int8 modes, which potentially offer two and eight times the fp32 rate respectively.

\section*{Acknowledgement}
For the purpose of open access, the author has applied a Creative Commons Attribution (CC BY) licence to any Author Accepted Manuscript version arising from this submission.


\bibliographystyle{IEEEtran}
\bibliography{references}
\end{document}